\documentclass[fleqn,11pt,twoside]{article}
\usepackage{amsthm,amsthm,amssymb, color, epsfig, graphics, subfigure,}
\usepackage{tikz}

\usepackage{amsmath}
\usepackage{graphicx}
\usepackage{amsmath}
\usepackage{graphicx}
\usepackage{latexsym}
\usepackage[all]{xy}
\usepackage{amsthm}
\usepackage{amsmath, amscd}
\usepackage{graphicx}
\usepackage{lscape}
\usepackage{amsmath}
\usepackage{graphicx}
\usepackage{latexsym}

\newtheorem{proposition}{Proposition}

\makeatletter
\newcommand{\copyrightnote}[2]{{\renewcommand{\thefootnote}{}
 \footnotetext{\small\it
\begin{flushleft}
Copyright \copyright \ #1 by  #2
\end{flushleft}}}}

\newcommand{\Name}[1]{\begin{flushleft}
                       \LARGE \bf #1
                       \end{flushleft}\vspace{-3mm}}

\newcommand{\Author}[1]{\begin{flushleft}
                       \it #1 \end{flushleft}}

\newcommand{\Address}[1]{\begin{flushleft}
                       \it #1 \end{flushleft}}

\newcommand{\Date}[1]{\begin{flushleft}
                      \small  \it #1 \end{flushleft}}

\newcommand{\evenhead}{Author \ name}
\newcommand{\oddhead}{Article \ name}

\renewcommand{\@evenhead}{
\hspace*{-3pt}\raisebox{-15pt}[\headheight][0pt]{\vbox{\hbox to \textwidth
{\thepage \hfil \evenhead}\vskip4pt \hrule}}}
\renewcommand{\@oddhead}{
\hspace*{-3pt}\raisebox{-15pt}[\headheight][0pt]{\vbox{\hbox to \textwidth
{\oddhead \hfil \thepage}\vskip4pt\hrule}}}
\renewcommand{\@evenfoot}{}
\renewcommand{\@oddfoot}{}

\long\def\@makecaption#1#2{%
  \vskip\abovecaptionskip
  \sbox\@tempboxa{\small \textbf{#1.}\ \ #2}%
  \ifdim \wd\@tempboxa >\hsize
    {\small \textbf{#1.}\ \ #2}\par
  \else
    \global \@minipagefalse
    \hb@xt@\hsize{\hfil\box\@tempboxa\hfil}%
  \fi
  \vskip\belowcaptionskip}

\newcommand{\JNMPnumberwithin}[3][\arabic]{%
  \@ifundefined{c@#2}{\@nocounterr{#2}}{%
    \@ifundefined{c@#3}{\@nocnterr{#3}}{%
      \@addtoreset{#2}{#3}%
      \@xp\xdef\csname the#2\endcsname{%
        \@xp\@nx\csname the#3\endcsname .\@nx#1{#2}}}}%
}

\newcommand{\resetfootnoterule} {
  \renewcommand\footnoterule{%
  \kern-3\p@
  \hrule\@width.4\columnwidth
  \kern2.6\p@}
}

\renewcommand{\footnoterule}{}

\newcommand{\be}{\begin{equation}}
\newcommand{\ee}{\end{equation}}
\newcommand{\ba}{\hspace*{-5pt}\begin{array}}
\newcommand{\ea}{\end{array}}
\newcommand{\p}{\partial}

\makeatother

\numberwithin{equation}{section}
\theoremstyle{definition}

\theoremstyle{proposition}

\renewcommand{\ba}{\begin{array}}
\renewcommand{\ea}{\end{array}}
\newcommand{\beg}{\begin{eqnarray}}
\newcommand{\eeq}{\end{eqnarray}}
\newcommand{\bg}{\begin{eqnarray*}}

\newcommand{\ed}{\end{eqnarray*}}
\newcommand{\n}{\newline\hfill}
\newcommand{\nn}{\nonumber}

\renewcommand{\p}{\partial} 
 
\newcommand{\notlhd}{\lhd\kern-.8em{/}\ } 
\newcommand{\notexist}{\ \exists\kern-.5em{\raise.1em\hbox{/}}\ }

\newcommand{\pde}[2]{\frac{\p #1}{\p #2}} 
 
\newcommand{\pdd}[2]{\frac{\p^2 #1}{\p #2^2}} 
\newcommand{\inp}{{\mbox{\vbox{\hrule width0ex\hbox{\vrule
 height0ex\kern3.8pt
\vbox{\kern2.5pt}\kern3.8pt \vrule height1.6ex}
\hrule width1.6ex}}}}

\makeatletter
\newcommand*{\da@rightarrow}{\mathchar"0\hexnumber@\symAMSa 4B }
\newcommand*{\da@leftarrow}{\mathchar"0\hexnumber@\symAMSa 4C }
\newcommand*{\xdashrightarrow}[2][]{%
  \mathrel{%
    \mathpalette{\da@xarrow{#1}{#2}{}\da@rightarrow{\,}{}}{}%
  }%
}
\newcommand{\xdashleftarrow}[2][]{%
  \mathrel{%
    \mathpalette{\da@xarrow{#1}{#2}\da@leftarrow{}{}{\,}}{}%
  }%
}
\newcommand*{\da@xarrow}[7]{%
  \sbox0{$\ifx#7\scriptstyle\scriptscriptstyle\else\scriptstyle\fi#5#1#6\m@th$}%
  \sbox2{$\ifx#7\scriptstyle\scriptscriptstyle\else\scriptstyle\fi#5#2#6\m@th$}%
  \sbox4{$#7\dabar@\m@th$}%
  \dimen@=\wd0 %
  \ifdim\wd2 >\dimen@
    \dimen@=\wd2 %
  \fi
  \count@=2 %
  \def\da@bars{\dabar@\dabar@}%
  \@whiledim\count@\wd4<\dimen@\do{%
    \advance\count@\@ne
    \expandafter\def\expandafter\da@bars\expandafter{%
      \da@bars
      \dabar@ 
    }%
  }%
  \mathrel{#3}%
  \mathrel{%
    \mathop{\da@bars}\limits
    \ifx\\#1\\%
    \else
      _{\copy0}%
    \fi
    \ifx\\#2\\%
    \else
      ^{\copy2}%
    \fi
  }%
  \mathrel{#4}%
}
\makeatother

\begin{document}


\renewcommand{\evenhead}{ {\LARGE\textcolor{blue!10!black!40!green}{{\sf \ \ \ ]ocnmp[}}}\strut\hfill 
M Euler and N Euler
}
\renewcommand{\oddhead}{ {\LARGE\textcolor{blue!10!black!40!green}{{\sf ]ocnmp[}}}\ \ \ \ \  
A link between 2nd-order and 3rd-order fully-nonlinear PDEs 
}


\thispagestyle{empty}
\newcommand{\FistPageHead}[3]{
\begin{flushleft}
\raisebox{8mm}[0pt][0pt]
{\footnotesize \sf
\parbox{150mm}{{Open Communications in Nonlinear Mathematical Physics}\ {\LARGE\textcolor{blue!10!black!40!green}{]ocnmp[}}
Special Issue 2, 2024\  pp
#2\hfill {\sc #3}}}\vspace{-13mm}
\end{flushleft}}

\FistPageHead{1}{\pageref{firstpage}--\pageref{lastpage}}{ \ \ }

\strut\hfill

\strut\hfill

\copyrightnote{The author(s). Distributed under a Creative Commons Attribution 4.0 International License}

\begin{center}

{\bf {\large Proceedings of the OCNMP-2024 Conference:\\ 

\smallskip

Bad Ems, 23-29 June 2024}}
\end{center}

\smallskip



\Name{On 2nd-order fully-nonlinear equations with links to 3rd-order fully-nonlinear equations}

\label{firstpage}




\Author{Marianna Euler and Norbert Euler}



\Address{
International Society of Nonlinear Mathematical Physics, 
Auf der Hardt 27,\\
56130 Bad Ems, Germany\\
and\\
Centro Internacional de Ciencias, Av. Universidad s/n, Colonia Chamilpa,\\
 62210 Cuernavaca, Morelos, Mexico\\
}


\Date{Received June 12, 2024; Accepted July 29, 2024}




\noindent
{\bf Abstract}: We derive the general conditions for fully-nonlinear symmetry-integrable second-order evolution equations and their first-order recursion operators. We then make us of the established Propositions to find a link between a class of fully-nonlinear third-order symmetry integrable evolution equations and fully-nonlinear second-order symmetry-integrable evolution equations.

\strut\hfill



\renewcommand{\theequation}{\arabic{section}.\arabic{equation}}

\allowdisplaybreaks

\section{Introduction}
We recently reported a class of third-order fully-nonlinear symmetry-integrable evolution equations in 1+1 dimensions
with rational nonlinearities in their highest derivative \cite{E-E-OCNMP-Dec2022}. For this class of equations we have furthermore reported all the potentialisations and some multi-potentialisations in \cite{EulerEuler2024}. It is interesting to note that some of these potential equations are in fact members of hierarchies of fully-nonlinear second-order evolution equations. In the current article we identify all those potential equations and give the explicit recursion operators that generate those hierarchies.
In particular, the fully-nonlinear third-order symmetry-integrable evolution equations that are relevant here are
\begin{subequations}
\begin{gather*}
u_t=\frac{u_{xx}^3}{u_{xxx}^2}\left(\lambda_1+\lambda_2 u_{xx}\right)^3
\end{gather*}
and
\begin{gather*}
u_t=\frac{1}{u_{xxx}^2},
\end{gather*}
\end{subequations}
where $\lambda_1$ and $\lambda_2$ are arbitrary constants.  For more details on symmetry-integrable hierarchies and recursion operators, we refer to \cite{Fokas} and \cite{E-E-OCNMP-Dec2022}, and the references therein. 

This article is organised as follows: In Section 2 we provide two Propositions by which one are able to identify 
second-order fully nonlinear equations that admit third-order Lie-Bäcklund symmetries (see Proposition 1) and by which we can obtain a recursion operator for those second-order fully-nonlinear equations (see Proposition 2). In Section 3 we apply the mentioned Propositions to establish links to the two fully-nonlinear equations given above. Finally in Section 4 we make our conclusions and propose some further studies that could be of interest.


\section{On fully-nonlinear second-order equations}

\noindent
We consider a second-order equation of the form
\begin{gather}
\label{F-general}
u_t=F(u,u_x,u_{xx})
\end{gather}
where the non-constant function $F$ is to be determined such that (\ref{F-general}) is fully nonlinear in $u_{xx}$ and admits a third-order Lie-Bäcklund symmetry generated by $\displaystyle{Z=\eta[u]\pde{\ }{u}}$; hence a recursion operator $R$ that generates this third-order symmetry and consequently an infinite number of higher-order Lie-Bäcklund symmetries. The hierarchy of symmetry-integrable evolution equations is then
\begin{gather}
u_{t_m}=R^{m}[u]\, u_t,\qquad m=0,1,2,\ldots\ .
\end{gather}
Our first step is to find the general form of $F$ and $\eta$ to satisfy this requirement as a necessary condition. This is given by

\strut\hfill

\begin{proposition}
\label{Prop-Gen-LB-3rd-order}
Consider the class of second-order evolution equations of the form (\ref{F-general}) viz
\begin{gather*}
u_t=F(u,u_x,u_{xx}).
\end{gather*}
Under the assumption that (\ref{F-general}) is symmetry-integrable, the following statements must be true:
\begin{enumerate}
\item
The function $F$ in  (\ref{F-general}) satisfies the condition
\begin{gather}
\label{SW-F}
\frac{\p^3F}{\p u_{xx}^3}
\left(\pde{F}{u_{xx}}\right)^{-1}
-\frac{3}{2}\left(\pdd{F}{u_{xx}}\right)^2
\left(\pde{F}{u_{xx}}\right)^{-2}=0
\end{gather}
with general solution
\begin{gather}
\label{F-ABC}
F(u,u_x,u_{xx})=-\frac{A(u,u_x)}{u_{xx}+B(u,u_x)}+C(u,u_x),
\end{gather}
where $A,\ B$ and $C$ are arbitrary functions of their arguments. The quasi-linear and semi-linear form of 
(\ref{F-general}) follow from the singular solution of (\ref{SW-F}), namely the solution for which
$\p^2F/\p u_{xx}^2=0$. 
[Note that (\ref{SW-F}) is the Schwarzian derivative in $F$.]

\item
Any third-order Lie-Bäcklund symmetry generator 
\begin{gather}
\label{LB-Gen3}
Z:=\eta(x,u,u_x,u_{xx},u_{xxx})\pde{\ }{u}
\end{gather}
for (\ref{F-ABC}) is of the general form
\begin{gather}
\eta(x,u,u_x,u_{xx},u_{xxx})=
\frac{c_0A^{3/2}u_{xxx}}{(u_{xx}+B)^3}
-\frac{c_0A^{3/2}}{(u_{xx}+B)^3}\left(B\pde{B}{u_x}-u_x\pde{B}{u}\right)\nn\\[0.3cm]
\qquad
+\frac{3c_0A^{1/2}}{2(u_{xx}+B)^2}\left(B\pde{A}{u_x}-u_x\pde{A}{u}\right)
-\frac{f_1}{u_{xx}+B}
+f_2,
\end{gather}
whereby $c_0$ denotes an arbitrary but non-zero constant, and the functions $A=A(u,u_x)$, $B=B(u,u_x)$, $C=C(u,u_x)$, $f_1=f_1(x,u,u_x)$ and $f_2=f_2(x,u,u_x)$ need to be determined such that the invariance condition
\begin{gather}
\left.
\vphantom{\frac{DA}{DB}}
L_E\eta (x,u,u_x,u_{xx},u_{xxx})\right|_{E=0}=0
\end{gather}
is satisfied. Here $E:=u_t-F(u,u_x,u_{xx})$ and
$L_E$ denotes the linear operator
\begin{gather}
L_E[u]:=\pde{E}{u}+\pde{E}{u_t}D_t+\pde{E}{u_x}D_x+\pde{E}{u_{xx}}D_x^2+\pde{E}{u_{xxx}}D_x^3.
\end{gather}

\end{enumerate}

\end{proposition}


We now seek 1st-order recursion operators $R$ for every equation that admits a third-order Lie-Bäcklund symmetry, which, according to Proposition \ref{Prop-Gen-LB-3rd-order}, are equations of the general form (\ref{F-ABC})
We consider both differential recursion operators
\begin{gather}
\label{Gen-R-Diff-2}
R[u]=G_1(u,u_x,u_{xx})D_x+G_0(u,u_x)
\end{gather}
as well as integro-differential recursion operators 
\begin{gather}
\label{R-Int-Gen}
R[u]=G_1(u,u_x,u_{xx})D_x+G_0(u,u_x)+I(u,u_x,u_{xx})D_x^{-1}\circ \Lambda(u,u_x,u_{xx}).
\end{gather}
Applying the standard condition for $R$ (see for example
\cite{Euler-Euler-book-article}), namely
\begin{gather}
\left.\vphantom{\frac{DA}{DB}}
\left[L_E,\ R[u]\right]=D_t R\right|_{E=0}
\end{gather}
we obtain the following


\strut\hfill

\begin{proposition}
\label{Prop-R-Diff-2}
For finding recursion operators of (\ref{F-ABC}) viz.
\begin{gather*}
u_t=-\frac{A(u,u_x)}{u_{xx}+B(u,u_x)}+C(u,u_x).
\end{gather*}
we distinguish between the following four cases:
\begin{enumerate}

\item
With the assumption $A=F_1(u_x)$ and $B(u,u_x)=0$ in (\ref{F-ABC}), it follows that
\begin{gather}
u_t=-\frac{F_1(u_x)}{u_{xx}}
+u\left(
\frac{a_1\sqrt{F_1(u_x)}}{u_x^2P(u_x)}
-\frac{F_1(u_x)}{u_xP(u_x)}\frac{dP}{du_x}
+\frac{F_1(u_x)}{u_x^2}
-\frac{1}{2u_x}\frac{dF_1}{du_x}
\right)\nn\\[0.3cm]
\qquad
+\frac{a_0\sqrt{F_1(u_x)}}{P(u_x)}+H(u_x)
\end{gather}
admits a recursion operator of the form (\ref{R-Int-Gen}) with
\begin{subequations}
\begin{gather}
G_1(u_x,u_{xx})=
\frac{k_1\sqrt{F_1(u_x)}}{u_{xx}}\\[0.3cm]
G_0(u_x)=
\frac{k_1\sqrt{F_1(u_x)}}{2P(u_x)}\frac{dP}{du_x}
-\frac{k_1\sqrt{F_1(u_x)}}{2u_x}
-\frac{a_1k_1}{2u_xP(u_x)}
+k_0\\[0.3cm]
I(u_x)=\alpha u_x\\[0.3cm]
\Lambda(u_x,u_{xx})=\frac{P(u_x)u_{xx}}{\sqrt{F_1(u_x)}},
\end{gather}
\end{subequations}
where $a_0,\ a_1$ and  $k_0$ are arbitrary constants, $k_1$ and $\alpha$ are an arbitrary non-zero constants, $H$ is an arbitrary function of $u_x$ and, furthermore, $F_1$ and $P$ must satisfy the following condition:
\begin{gather}
0=k_1u_x^2P^2\left(u_x\frac{dP}{du_x}+P\right)\frac{d^2F_1}{du_x^2}
-k_1u_xP\left[
-3u_x^2P\frac{d^2P}{du_x^2}
+2u_x^2\left(\frac{dP}{du_x}\right)^2
-2u_xP\frac{dP}{du_x}
\right.\nn\\[0.3cm]
\left.
\vphantom{\frac{DA}{DB}}
+2P^2\right]\frac{dF_1}{du_x}
+2k_1\left[
u_x^3P^2\frac{d^3P}{du_x^3}
-2u_x^3P\frac{dP}{du_x}  \frac{d^2P}{du_x^2}
+u_x^2P^2\frac{d^2P}{du_x^2}
+u_x^3\left(\frac{dP}{du_x}\right)^3
\right.\nn\\[0.3cm]
\left.
-u_x^2P\left(\frac{dP}{du_x}\right)^2
-u_xP^2\frac{dP}{du_x}
+P^3\right]F_1
+2\left(4\alpha u_x^3P^3-k_1a_1^2\right)\left(u_x\frac{dP}{du_x}+P\right).
\end{gather}

\item
With the assumption $A=F_1(u_x)$ and $B=B(u_x)$  in (\ref{F-ABC}), it follows that
\begin{gather}
u_t=-\frac{F_1(u_x)}{u_{xx}+(b_1+b_2u_x)\sqrt{F_1(u_x)} }
+
\frac{\sqrt{F_1(u_x)}}{b_1+b_2u_x}
\end{gather}
admits a recursion operator of the form (\ref{R-Int-Gen}) with
\begin{subequations}
\begin{gather}
G_1(u_x,u_{xx})=
\frac{k_1\sqrt{F_1(u_x)}}{u_{xx}+(b_1+b_2u_x)\sqrt{F_1(u_x)} }\\[0.3cm]
G_0(u_x)=
\frac{k_1b_2F_1(u_x)}{u_{xx}+(b_1+b_2u_x)\sqrt{F_1(u_x)}}
-\frac{k_1b_2\sqrt{F_1(u_x)}}{b_1+b_2u_x}+k_0\\[0.3cm]
I(u_x)= u_x+\frac{b_1}{b_2}\\[0.3cm]
\Lambda(u_x,u_{xx})=\alpha b_2\left(\frac{u_{xx}+(b_1+b_2u_x)
\sqrt{F_1(u_x)}}{(b_1+b_2u_x)\sqrt{F_1(u_x)}}\right),
\end{gather}
\end{subequations}
where $b_1$ and $\alpha$ are arbitrary constants, whereas $b_2$ and $k_1$ are arbitrary non-zero constants.

\item
With the assumption $A=F_1(u_x)$, $B=0$ and $C=C(u_x)$  in (\ref{F-ABC}), it follows that
\begin{gather}
\label{EQ-F1-C}
u_t=-\frac{F_1(u_x)}{u_{xx}}+C(u_x).
\end{gather}
This leads to two subcases:

\smallskip

\noindent
3.1. Equation (\ref{EQ-F1-C}) admits the recursion operator of the form (\ref{R-Int-Gen}) with
\begin{subequations}
\begin{gather}
G_1(u_x,u_{xx})=
\frac{k_1\sqrt{F_1(u_x)}}{u_{xx} }\\[0.3cm]
G_0(u_x)=
-\frac{k_1}{4\sqrt{F_1(u_x)}}\frac{dF_1}{du_x}
+k_0\\[0.3cm]
I(u_x)= \alpha \\[0.3cm]
\Lambda(u_x,u_{xx})=\frac{u_{xx}}{F_1(u_x)},
\end{gather}
\end{subequations}
where $k_0$ is an arbitrary constant, $k_1$ and $\alpha$ are arbitrary non-zero constants, 
$C$ is an arbitrary function of $u_x$,
and $F_1$ must satisfy the following condition:
\begin{gather}
k_1\left(4F_1^4\frac{d^3 F_1}{du_x^3}
-6F_1^3\frac{dF_1}{du_x}\frac{d^2F_1}{du_x^2}
+3F_1^2\left(\frac{dF_1}{du_x}\right)^3\right)
+16\alpha F_1^{5/2}\frac{dF_1}{du_x}=0.
\end{gather}

\noindent
3.2. Equation (\ref{EQ-F1-C}) admits the recursion operator of the form (\ref{R-Int-Gen}) with
\begin{subequations}
\begin{gather}
G_1(u_x,u_{xx})=
\frac{k_1\sqrt{F_1(u_x)}}{u_{xx} }\\[0.3cm]
G_0(u_x)=
-\frac{k_1}{4\sqrt{F_1(u_x)}}\frac{dF_1}{du_x}
+k_0\\[0.3cm]
I(u_x)= \alpha u_x\\[0.3cm]
\Lambda(u_x,u_{xx})=\frac{u_x u_{xx}}{F_1(u_x)},
\end{gather}
\end{subequations}
where $k_0$ is an arbitrary constant, $k_1$ and $\alpha$ are arbitrary non-zero constants,
$C$ is an arbitrary function of $u_x$,
and $F_1$ must satisfy the following condition:
\begin{gather}
k_1\left(4F_1^4\frac{d^3 F_1}{du_x^3}
-6F_1^3\frac{dF_1}{du_x}\frac{d^2F_1}{du_x^2}
+3F_1^2\left(\frac{dF_1}{du_x}\right)^3\right)\nn\\[0.3cm]
+16\alpha F_1^{5/2}u_x\left(u_x\frac{dF_1}{du_x}-4F_1\right)=0.
\end{gather}

\item
The equation
\begin{gather}
\label{Eq-F1-F2}
u_t=-\frac{F_1(u_x)}{u_{xx}+(b_1+b_2u_x)\sqrt{F_1(u_x)}}+F_2(u_x)
\end{gather}
admits a recursion operator of the form (\ref{Gen-R-Diff-2}) with
\begin{subequations}
\begin{gather}
G_1(u_x)=\frac{k_1\sqrt{F_1}}{u_{xx}+(b_1+u_xb_2)\sqrt{F_1}}\\[0.3cm]
G_0(u_x)=
\frac{k_1b_2F_1}{u_{xx}+(b_1+b_2 u_x)\sqrt{F_1}}
-\frac{k_1}{4\sqrt{F_1}}\frac{dF_1}{du_x}
+\frac{k_1}{2}(b_1+b_2 u_x)\frac{dF_2}{du_x}\nn\\[0.3cm]
\qquad
-\frac{k_1b_2}{2}F_2
+k_0,
\end{gather}
\end{subequations}
where $b_1$ and $b_2$ are arbitrary constants, $k_1$ is an arbitrary non-zero constant, and 
$F_1$ and $F_2$ must satisfy the following condition:
\begin{gather}
4F_1^2\frac{d^3 F_1}{du_x^3}
-6F_1\frac{dF_1}{du_x}\frac{d^2F_1}{du_x^2}
+3\left(\frac{dF_1}{du_x}\right)^3
-8F_1^{5/2} \left(\left(b_1+b_2u_x\right) \frac{d^3F_2}{du_x^3}
\right.
\nn\\[0.3cm]
\left.
+3b_2\frac{d^2F_2}{du_x^2}\right)
=0.
\end{gather}

\end{enumerate}

\end{proposition}

\section{Fully-nonlinear second-order equations related to fully-non-\n
linear third-order equations}

We now apply Proposition 1 and Proposition 2 to identify fully-nonlinear second-order equations that are related to fully-nonlinear third-order equations. For this purpose we consider two main cases:

\strut\hfill

\noindent
{\bf Case 1:} Consider the fully-nonlinear symmetry-integrable equation \cite{E-E-OCNMP-Dec2022}
\begin{gather}
\label{FN-3rd-order-G1}
u_t=\frac{u_{xx}^3}{u_{xxx}^2}\left(\lambda_1+\lambda_2 u_{xx}\right)^3,
\end{gather}
where $\lambda_1$ and $\lambda_2$ are arbitrary constants. For the potentialisations of this equation we need to 
distinguish different subcases that depend on the constants $\lambda_1 $ and $\lambda_2$.

\strut\hfill

\noindent
{\bf Subcase 1.1:} Consider equation (\ref{FN-3rd-order-G1}), with $\lambda_1\neq 0$ and $\lambda_2\neq 0$.
Then (\ref{FN-3rd-order-G1}) admits the zero-order potentialisation \cite{EulerEuler2024} 
\begin{gather}
\label{vt-3rd-case11}
v_t=
\frac{\lambda_1^3}{4}
\left(
\vphantom{\frac{A}{B}}
1+\lambda_1^2\lambda_2v_x^2
\right)^{3/2}
\left(
\frac{v_x^3v_{xxx}}{v_{xx}^3}
-\frac{3v_x^2}{v_{xx}}
\right),
\end{gather}
where 
\begin{gather}
v_x=-\frac{2}{\lambda_1^2}\left[u_{xx}(\lambda_1+\lambda_2u_{xx})\right]^{1/2}.
\end{gather}
Applying Proposition 1 we find that the fully-nonlinear second-order equation
\begin{gather}
\label{vt-2nd-case11}
\boxed{\vphantom{\frac{DA}{DB}}
v_t=- \left(
\vphantom{\frac{A}{B}}
1+\lambda_1^2\lambda_2 v_x^2\right)   \frac{v_x^2}{v_{xx}}-v
}
\end{gather}
admits the Lie-Bäcklund symmetry  generator $\displaystyle{Z=\eta[v]\pde{\ }{v}}$
 for which (\ref{vt-3rd-case11}) is the third-order flow, i.e.
\begin{gather}
\eta(x,v,v_x,v_{xx},v_{xxx})=\frac{\lambda_1^3}{4}
\left(
\vphantom{\frac{A}{B}}
1+\lambda_1^2\lambda_2v_x^2
\right)^{3/2}
\left(
\frac{v_x^3v_{xxx}}{v_{xx}^3}
-\frac{3v_x^2}{v_{xx}}
\right).
\end{gather}
Applying now Proposition 2 we find that equation (\ref{vt-2nd-case11}) admits a recursion operator $R_{11}[v]$ of the form 
(\ref{R-Int-Gen}),
where
\begin{subequations}
\begin{gather}
G_1=\frac{\lambda_1^3}{4}
\left(
\vphantom{\frac{A}{B}}
1+\lambda_1^2\lambda_2 v_x^2\right)^{1/2}
\frac{v_x}{v_{xx}}\\[0.3cm]
G_0=-\frac{\lambda_1^5\lambda_2}{4}
\left(
\vphantom{\frac{A}{B}}
1+\lambda_1^2\lambda_2 v_x^2\right)^{-1/2}v_x^2
\\[0.3cm]
I=v_x
\\[0.3cm]
\Lambda=\frac{\lambda_1^5\lambda_2}{4}
\left(
\vphantom{\frac{A}{B}}
1+\lambda_1^2\lambda_2 v_x^2\right)^{-3/2}v_{xx}
\end{gather}
\end{subequations}
for which (\ref{vt-3rd-case11}) is the second member $v_{t_1}$ in the hierarchy 
$\displaystyle{v_{t_m}=R_{11}^m[v]v_t}$, $m=0,1,2,\ldots$ and $v_t$ is the equation (\ref{vt-2nd-case11}) .

\strut\hfill

\noindent
{\bf Subcase 1.2:} Consider equation (\ref{FN-3rd-order-G1}), with $\lambda_1=0$ and $\lambda_2=1$, i.e.
\begin{gather}
\label{FN-3rd-order-case12}
u_t=\frac{u_{xx}^6}{u_{xxx}^2}
\end{gather}
Then (\ref{FN-3rd-order-case12}) admits the potentialisation \cite{EulerEuler2024}
\begin{gather}
\label{vt-3rd-case12}
v_t=
\frac{v_x^6v_{xxx}}{v_{xx}^3}
-\frac{3v_x^5}{v_{xx}},
\end{gather}
where 
\begin{gather}
v_x=-2^{1/3} u_{xx}.
\end{gather}
Applying Proposition 1 we find that the fully-nonlinear second-order equation
\begin{gather}
\label{vt-2nd-case12}
\boxed{\vphantom{\frac{DA}{DB}}
v_t=- \frac{v_x^4}{v_{xx}}
}
\end{gather}
admits the Lie-Bäcklund symmetry  generator $\displaystyle{Z=\eta[v]\pde{\ }{v}}$
 for which (\ref{vt-3rd-case12}) is the third-order flow, i.e.
\begin{gather}
\eta(x,v,v_x,v_{xx},v_{xxx})=
\frac{v_x^6v_{xxx}}{v_{xx}^3}
-\frac{3v_x^5}{v_{xx}}.
\end{gather}
Applying now Proposition 2 we find that equation (\ref{vt-2nd-case12}) admits a recursion operator $R_{12}[v]$ of the form (\ref{Gen-R-Diff-2}),
where
\begin{subequations}
\begin{gather}
G_1=
\frac{v_x^2}{v_{xx}}
\\[0.3cm]
G_0=-v_x
\end{gather}
\end{subequations}
for which (\ref{vt-3rd-case12}) is the second member $v_{t_1}$ in the hierarchy 
$\displaystyle{v_{t_m}=R_{12}^m[v]v_t}$, $m=0,1,2,\ldots$ and $v_t$ is the equation (\ref{vt-2nd-case12}) .

\strut\hfill

\noindent
{\bf Subcase 1.3:} Consider equation (\ref{FN-3rd-order-G1}), with $\lambda_1=-1$ and $\lambda_2=0$, i.e.
\begin{gather}
\label{FN-3rd-order-case13}
u_t=-\frac{u_{xx}^3}{u_{xxx}^2}
\end{gather}
For this equation we have established several potentialisations in \cite{EulerEuler2024}. We treat the relevant cases below:

\begin{itemize}
\item[{\bf 1.3a:}]
Equation (\ref{FN-3rd-order-case13}) admits the potentialisation \cite{EulerEuler2024}
\begin{gather}
\label{vt-3rd-case13}
v_t=
\frac{2v_x^3v_{xxx}}{v_{xx}^3}
-\frac{3v_x^2}{v_{xx}},
\end{gather}
where 
\begin{gather}
v_x=-u_{xx}.
\end{gather}
Applying Proposition 1 we find that the fully-nonlinear second-order equation
\begin{gather}
\label{vt-2nd-case13}
\boxed{\vphantom{\frac{DA}{DB}}
v_t=- \frac{v_x^2}{v_{xx}}
}
\end{gather}
admits the Lie-Bäcklund symmetry generator $\displaystyle{Z=\eta[v]\pde{\ }{v}}$
 for which (\ref{vt-3rd-case13}) is the third-order flow, i.e.
\begin{gather}
\eta(x,v,v_x,v_{xx},v_{xxx})=
\frac{2v_x^3v_{xxx}}{v_{xx}^3}
-\frac{3v_x^2}{v_{xx}}.
\end{gather}
Applying now Proposition 2 we find that equation (\ref{vt-2nd-case13}) admits a recursion operator $R_{13}[v]$ of the form (\ref{Gen-R-Diff-2}),
where
\begin{subequations}
\begin{gather}
G_1=
\frac{2v_x}{v_{xx}}
\\[0.3cm]
G_0=-1
\end{gather}
\end{subequations}
for which (\ref{vt-3rd-case13}) is the second member $v_{t_1}$ in the hierarchy 
$\displaystyle{v_{t_m}=R_{13}^m[v]v_t}$, $m=0,1,2,\ldots$ and $v_t$ is the equation (\ref{vt-2nd-case13}).

\item[{\bf 1.3b:}]
Furthermore we know from the results reported in \cite{EulerEuler2024} that a zero-order potentialisation of equation (\ref{vt-3rd-case13}) is 
\begin{gather}
\label{vt-3rd-case13-add1}
V_t=\frac{V_{xxx}}{V_{xx}^3}-3\cdot 2^{-2/3}\frac{1}{V_{xx}}-2^{-1/3}\,x,
\end{gather}
where 
\begin{gather}
V_x=2^{-1/2}\ln(v_x)
\end{gather}
gives the relation to (\ref{vt-3rd-case13}) and
\begin{gather}
V_x=2^{-1/2}\ln|-u_{xx}|
\end{gather}
the relation to (\ref{FN-3rd-order-case13}).
Applying Proposition 1 for (\ref{vt-3rd-case13-add1})
we find that the fully-nonlinear second-order equation
\begin{gather}
\label{vt-2nd-case13-add1}
\boxed{\vphantom{\frac{DA}{DB}}
V_t=- \frac{1}{V_{xx}}
}
\end{gather}
admits the Lie-Bäcklund symmetry generator $\displaystyle{Z=\eta[V]\pde{\ }{V}}$ for which (\ref{vt-3rd-case13-add1}) is the third-order flow, i.e.
\begin{gather}
\eta(x,V,V_x,V_{xx},V_{xxx})=
\frac{V_{xxx}}{V_{xx}^3}-3\cdot 2^{-2/3}\frac{1}{V_{xx}}-2^{-1/3}\,x .
\end{gather}
Applying now Proposition 2 we find that equation (\ref{vt-2nd-case13-add1}) admits a recursion operator $R_{131}[V]$ of the form (\ref{R-Int-Gen}),
where
\begin{subequations}
\begin{gather}
G_1=\frac{1}{V_{xx}}
\\[0.3cm]
G_0=3\cdot 2^{-2/3}
\\[0.3cm]
I=1
\\[0.3cm]
\Lambda=2^{-1/3}V_{xx}
\end{gather}
\end{subequations}
for which (\ref{vt-3rd-case13-add1}) is the second member $V_{t_1}$ in the hierarchy 
$\displaystyle{V_{t_m}=R_{131}^m[V]V_t}$, $m=0,1,2,\ldots$ and $V_t$ is the equation (\ref{vt-2nd-case13-add1}).

\item[{\bf 1.3c:}]
By a multi-potentialisations of (\ref{FN-3rd-order-case13}) we are led to the equation \cite{EulerEuler2024}
\begin{gather}
\label{vt-3rd-case13-add2}
w_{t}=\frac{w_{xxx}}{w_{xx}^3}-2^{-2/3}\frac{3}{w_{xx}},
\end{gather}
where
\begin{gather}
w_{xx}=2^{-1/3}\frac{u_{xxx}}{u_{xx}}
\end{gather}
gives the relation to (\ref{FN-3rd-order-case13}).
Applying Proposition 1 for (\ref{vt-3rd-case13-add2})
we find that the fully-nonlinear second-order equation
\begin{gather}
\label{vt-2nd-case13-add2}
\boxed{\vphantom{\frac{DA}{DB}}
w_{t}=- \frac{1}{w_{xx}}
}
\end{gather}
admits the Lie-Bäcklund symmetry generator $\displaystyle{Z=\eta[w]\pde{\ }{w}}$ for which (\ref{vt-3rd-case13-add2}) is the third-order flow, i.e.
\begin{gather}
\eta(x,w,w_{x},w_{xx},w_{xxx})=
\frac{w_{xxx}}{w_{xx}^3}-2^{-2/3}\frac{3}{w_{xx}}.
\end{gather}
Applying now Proposition 2 we find that equation (\ref{vt-2nd-case13-add2}) admits a recursion operator $R_{132}[w]$ of the form (\ref{Gen-R-Diff-2}),
where
\begin{subequations}
\begin{gather}
G_1=\frac{1}{w_{xx}}
\\[0.3cm]
G_0=3\cdot 2^{-2/3}
\end{gather}
\end{subequations}
for which (\ref{vt-3rd-case13-add2}) is the second member $w_{t_1}$ in the hierarchy 
$\displaystyle{w_{t_m}=R_{132}^m[w]w_{t}}$, $m=0,1,2,\ldots$ and $w_{t}$ is the equation (\ref{vt-2nd-case13-add2}).

\end{itemize}

\strut\hfill

\noindent
{\bf Case 2:} Consider the fully-nonlinear symmetry-integrable equation \cite{E-E-OCNMP-Dec2022}
\begin{gather}
\label{FN-3rd-order-G2}
u_t=\frac{1}{u_{xxx}^2}.
\end{gather}
A multi-potentialisation of (\ref{FN-3rd-order-G2}) leads the quasi-linear equation \cite{EulerEuler2024}
\begin{gather}
\label{vt-3rd-case14}
v_t=\frac{v_{xxx}}{v_{xx}^3},
\end{gather}
where 
\begin{gather}
v_x=-2^{-1/3}u_{xx}.
\end{gather}
Applying Proposition 1 we find that the fully-nonlinear second-order equation
\begin{gather}
\label{vt-2nd-case14}
\boxed{\vphantom{\frac{DA}{DB}}
v_t=- \frac{1}{v_{xx}}
}
\end{gather}
admits the Lie-Bäcklund symmetry generator $\displaystyle{Z=\eta[v]\pde{\ }{v}}$
 for which (\ref{vt-3rd-case14}) is the third-order flow, i.e.
\begin{gather}
\eta(x,v,v_x,v_{xx},v_{xxx})=
\frac{v_{xxx}}{v_{xx}^3}.
\end{gather}
Applying now Proposition 2 we find that equation (\ref{vt-2nd-case14}) admits a recursion operator $R_{14}[v]$ of the form (\ref{Gen-R-Diff-2}),
where
\begin{subequations}
\begin{gather}
G_1=
-\frac{1}{v_{xx}}
\\[0.3cm]
G_0=0
\end{gather}
\end{subequations}
for which (\ref{vt-3rd-case14}) is the second member $v_{t_1}$ in the hierarchy 
$\displaystyle{v_{t_m}=R_{14}^m[v]v_t}$, $m=0,1,2,\ldots$ and $v_t$ is the equation (\ref{vt-2nd-case14}).

\section{Concluding remarks}

In this article we have provided examples of third-order fully-nonlinear symmetry-integrable equations that are linked to 
second-order fully nonlinear symmetrey-integrable equations. In particular, the examples of Case 1 and Case 2 show that certain potentialisations of considered third-order fully-nonlinear equations are in fact the third-order flows of the Lie-Bäcklund symmetries of particular fully-nonlinear second-order equations. Hence these potential equations, which are quasi-linear third-order equations, are members of hierarchies of symmetry-integrable equations for which the seed equation is a second-order fully-nonlinear equation.

We should point out that not every fully-nonlinear third-order symmetry-integrable equation can be linked in this way to a second-order equation. Moreover, not every potential equation that results from third-order fully-nonlinear symmetry-integrable equation belongs to a symmetry-integrable hierarchy of second-order equations, even if it does so for other potentialisations. For example, the fully-nonlinear third-order symmetry integrable equation (\ref{FN-3rd-order-case12}), viz
\begin{gather*}
u_t=\frac{u_{xx}^6}{u_{xxx}^2}
\end{gather*}
is related to the fully-nonlinear second-order equation 
\begin{gather*}
v_t=-\frac{v_x^4}{v_{xx}}
\end{gather*}
via the potential equation
\begin{gather*}
v_t=\frac{v_x^6v_{xxx}}{v_{xx}^3}-\frac{3v_x^5}{v_{xx}}
\end{gather*}
as shown in Subcase 1.2. However, as reported in \cite{EulerEuler2024},
equation (\ref{FN-3rd-order-case12}) also admits the zero-order potentialisation 
\begin{gather}
\label{counter-ex}
v_t=\frac{v_{xxx}}{v_{xx}^3}+\frac{3}{v_xv_{xx}},
\end{gather}
where 
\begin{gather*}
v_x=\frac{1}{2^{1/3}u_{xx}}.
\end{gather*}
Using Proposition 1 it is easy to shown that there exists no second-order equation that admits the Lie-Bäcklund symmetry generator
\begin{gather*}
Z=\left(\frac{v_{xxx}}{v_{xx}^3}+\frac{3}{v_xv_{xx}}  \right)\pde{\ }{v}.
\end{gather*}
We conclude that the potential equation (\ref{counter-ex}) does not provide a link between a second-order equation and the third-order fully-nonlinear equation (\ref{FN-3rd-order-case12}). Furthermore, and example of a fully-nonlinear third-order symmetry-integrable equation that does not admit a potentialisation that links this equation to a second-order symmetry-integrable equation is \cite{EulerEuler2024}
\begin{gather*}
u_t=\frac{4u_x^5}{(2bu_x^2-2u_xu_{xxx}+3u_{xx}^2)^2},
\end{gather*}
for any constant $b$.

Finally, we give an example of a more general second-order fully-nonlinear symmetry-integrable evolution than those that are known to use to be linked to a fully-nonlinear third-order equation from the results reported in \cite{EulerEuler2024}: Applying Proposition 1 and Proposition 2 we find that
\begin{gather}
\label{L-Ex}
v_t=-\frac{v_x^2}{v_{xx}}+c_3v+c_4v_x\ln(v_x)+c_4+c_5v_x
\end{gather}
admits the recursion operator
\begin{gather}
R[v]=\frac{27}{4}\left(\frac{v_x}{v_{xx}}\right)D_x-\frac{27}{4}+\frac{27c_3}{4}+\alpha v_xD_x^{-1}\circ \frac{v_{xx}}{v_x^2},
\end{gather}
where $\alpha$, $c_3,\ c_4$ and $c_5$ are arbitrary constants. 
The third-order equation in the hierarchy $v_{t_m}=R^m[v]v_t$ is then
\begin{gather}
\label{vt-27-g}
v_{t_1}=\frac{27}{4}\frac{v_x^3v_{xxx}}{v_{xx}^3}
-\frac{27}{4}\frac{v_x^2}{v_{xx}}
+\alpha (c_3-1)x v_x
+c_3\left(\frac{27c_3}{4}-\frac{27}{4}-\alpha\right)v
+\frac{27}{4}(c_4+c_3c_5)v_x\nn\\[0.3cm]
\qquad
+\left(\frac{27c_3c_4}{4}+\alpha c_5\right)v_x\ln(v_x)
+\frac{\alpha c_4}{2}v_x\ln^2(v_x)
+c_4\left(\frac{27}{4}(c_3-1)-\alpha\right).
\end{gather}
By letting $c_3=c_4=c_5=\alpha=0$ we obtain 
\begin{gather}
\label{vt-27}
v_{t_1}=\frac{27}{4}\frac{v_x^3v_{xxx}}{v_{xx}^3}
-\frac{27}{4}\frac{v_x^2}{v_{xx}}
\end{gather}
which is an equation that was obtained in \cite{EulerEuler2024} by a multi-potentialisation of (\ref{FN-3rd-order-case13}), viz
\begin{gather*}
u_t=-\frac{u_{xx}^3}{u_{xxx}^2}.
\end{gather*}
The relation between (\ref{FN-3rd-order-case13}) and (\ref{vt-27}) is given my
\begin{gather}
\frac{v_{xx}}{v_x}=\frac{3}{2}\frac{u_{xxx}}{u_{xx}}.
\end{gather}
In view of this example one may ask whether this method could be used to construct further fully-nonlinear third-order symmetry-integrable evolution equations different from those that have been reported in \cite{E-E-OCNMP-Dec2022}. 
Furthermore, the method proposed here could be exploited for the construction of higher-order fully-nonlinear symmetry-integrable equations. 

\label{lastpage}


\begin{thebibliography} {99}


\bibitem{Euler-Euler-book-article}
 Euler M and Euler N, Nonlocal invariance of the multipotentialisations of the Kupershmidt equation and its higher-order hierarchies In:
  {\it Nonlinear Systems and Their Remarkable Mathematical Structures}, N Euler (ed), CRC Press, Boca Raton, 317--351, 2018.



\bibitem{E-E-OCNMP-Dec2022}
Euler M and Euler N, On fully-nonlinear symmetry-integrable equations
with rational functions in their highest derivative: Recursion operators,
{\it Open Communications in Nonlinear Mathematical Physics}, {\bf 2}, 216--228, 2022.


\bibitem{EulerEuler2024}
Euler M and Euler N,
Potentialisations of a class of fully-nonlinear symmetry-integrable evolution equations,
{\it Open Communications in Nonlinear Mathematical Physics}, {\bf 4}, 44--78, 2024.

  

\bibitem{Fokas}
Fokas A S and Fuchssteiner B, On the structure of symplectic operators and hereditary symmetries,
 {\it Lettere al Nuovo Cimento} {\bf 28}, 299--303, 1980.




\end {thebibliography}
\end{document}